\documentclass[RNAAS]{aastex63}

\submitjournal{RNAAS}

\shorttitle{Dissociation of H$_2$}
\shortauthors{Shaw, Ferland \& Ploeckinger}

\begin{document}

\title{Cosmic ray dissociation of molecular hydrogen and dense cloud chemistry}

\correspondingauthor{Gargi Shaw}
\email{gargishaw@gmail.com, gary@g.uky.edu}

\author{Gargi Shaw}
\affiliation{Department of Astronomy and Astrophysics, Tata Institute of Fundamental Research\\
Homi Bhabha Road, Navy Nagar, Colaba, Mumbai 400005, India}

\author{G. J. Ferland}
\affiliation{Department of Physics and Astronomy, University of Kentucky\\
Lexington, KY 40506, USA}

\author{S. Ploeckinger}
\affiliation{Lorentz Institute for theoretical physics, Leiden University\\
PO Box 9506, NL-2300 RA Leiden, the Netherlands}



\begin{abstract}
Dissociation of molecular hydrogen by secondary electrons produced by cosmic ray or X-ray ionization
plays a crucial role in the chemistry of the densest part of  molecular clouds. 
Here we study the effect of the mean kinetic energy of secondary electrons on this process.
We compare predictions using a range of secondary electron energies
and  predictions of the cross sections with the values in the UMIST database.
We find that the predicted  column densities change by nearly one dex.

\end{abstract}


\section{Introduction} \label{sec:intro}

Observationally, a significant fraction of atomic hydrogen (H) mixed with molecular hydrogen (H$_2$) is found to be present in the  densest part of 
clouds that are
shielded from the  UV flux \citep{2003ApJ...585..823L}. This is due to the dissociation of H$_2$ into atomic hydrogen by cosmic rays (CRs). 
The secondary electrons generated by CRs excite H$_2$ to its repulsive triplet state b that leads to dissociation of H$_2$ into two H atoms. 
The secondary electrons can also excite H$_2$ to the other triplet bound states e, a, and c, which will eventually decay down to the b state. 
In addition to this,
CRs ionize H$_2$ into H$_2$$^+$. \citet{2018A&A...619A.144P} showed that the ratio of dissociation rate to the ionization rate ($\it R_{diss/ion}$) 
varies between 0.6 to 0.7. 
The UMIST database \citep{1997A&AS..121..139M} mentions 1.3$\times$10$^{-18}$ s$^{-1}$ and  1.2$\times$10$^{-17}$ s$^{-1}$ as 
the dissociation and 
ionization rates, respectively. This leads to $\it R_{diss/ion}$ = 0.1 for UMIST rates. 
It is to be noted that the origin of the UMIST rates is not given.
The abundance of H has important consequences for the chemical evolution of dense regions in the clouds, 
and this is affected by these rates.

\citet{1999ApJS..125..237D} calculated the cross-sections for the H$_2$ dissociation as a function of 
the incident electron energy (their fig 4b). 
This cross-section is not uniform but rather decreases rapidly for higher energy. 
It is known that the secondary electrons have a broad range of energies. 
\citet{1968ApJ...152..971S} states that the secondary electrons have 
kinetic energies
between about 15 and 30 eV for inelastic collisions between 35 and 50 eV for elastic collisions. 
\citet{1958RSPSA.248..415D} and \citet{1973ApJ...186..859G} report that the mean energy of the 
secondary electrons can be
taken as 36 eV. 
Here we aim to study the effect of these different cross-sections at energies, 15 eV, 30 eV, and 36 eV on the 
cloud's chemistry and $\it R_{diss/ion}$ and compare this with the rate adopted by UMIST.

\section{Calculations and Results} \label{sec:results}

We consider a simple plane-parallel model with standard ISM abundances of gas and dust and set a constant 
kinetic temperature of 50 K.  
The hydrogen density  is 10$^5$ cm$^{-3}$. Here we consider cosmic rays as the only source of radiation, and we block all 
UV flux and photoionization. 
The cosmic-ray ionization rate of H varies along different sight lines \citep{2007ApJ...671.1736I}. 
For our model, we consider this to be 
10$^{-16}$ s$^{-1}$. The cloud thickness is A$_V$ = 10. 
All  models presented here are calculated using version 17.01 of the spectral simulation code, CLOUDY 
\citep{{2013RMxAA..49..137F},{2017RMxAA..53..385F},{2005ApJ...624..794S}}. 

We use  figure 4b in \citet{1999ApJS..125..237D} to evaluate the excitation cross section for 
the dissociative triplet state b \citep{2008ApJ...675..405S}, 
rescale it in terms of the hydrogen ionization crosssection \citep{1985ApJ...296..765S}, 
and then multiply by the cosmic-ray excitation rate of Ly$\alpha$ at secondary electron 
energies at the three energies, 15 eV, 30eV, and 36 eV.
For this simple model, the ratio $\it R_{diss/ion}$ is 9.93, 2.78, and 0.93 for cases with secondary electron energies at 15 eV, 30eV, and 36 eV, respectively.  

Table  \ref{tab:1} shows predicted column densities of the most-affected species  
 and also shows the results using UMIST rates for 
the above mentioned two reactions, keeping other reactions the  same. 
The UMIST rates predict the lowest H column density. 
Although the column densities in the lower levels of H$_2$ do 
not change very much, the column densities for higher levels (H$_2$$^*$) do change significantly. 
Note that the H$_3^+$ column density, the current gold standard for measuring the cosmic ray
background \citep{2007ApJ...671.1736I}, varies by one dex.


\begin{deluxetable}{ccccl}
\tablecaption{Predicted column densities (cm$^{-2}$) in log scale \label{tab:1}}
\tablehead{
\colhead{Species} & \colhead{15 eV} & \colhead{30 eV} & \colhead{36 eV} &\colhead{UMIST}
}
\startdata
H  & 18.73  & 18.32  & 18.09 & 17.28 \\
H$^+$ & 12.09  & 11.90 & 11.82 & 11.92 \\
OH & 11.70 & 11.52 & 11.45 & 10.70\\
LiH & 4.11 & 4.50 & 4.75 & 5.61 \\
C$_3$H$^+$ & 12.54 & 12.71 & 12.77 & 12.61 \\
C$_2$ & 17.09 & 17.23 & 17.28 & 16.47 \\
C$_3$ & 16.29 & 16.46 & 16.52 & 16.36\\
CN & 15.90 & 16.01 & 16.05 & 15.35\\
SiO & 13.22 & 13.05 & 12.99 & 12.55\\
HS & 13.85 & 13.67 & 13.61 & 13.54\\
NS & 14.61 & 14.39 & 14.31 & 13.94\\
SO & 11.25 & 10.86 & 10.72 & 10.92\\
H$_2$$^*$ & 13.96 & 13.54 & 13.31 & 12.51\\
H$_3$$^+$ & 13.86 & 13.85 & 13.84 & 12.68\\
\enddata
\end{deluxetable}  

Earlier, Cloudy has used 20 eV secondary electrons as described in \citet{2008ApJ...675..405S}.
In future versions of Cloudy, we will use 36 eV as the mean kinetic energy of the secondary electrons.  
The ideal calculation would solve for the cosmic ray secondary electron energy spectrum.

\acknowledgments

GS acknowledges WOS-A grant from Department of Science and Technology (SR/WOS-A/PM-9/2017). 
GJF acknowledges support by NSF (1816537, 1910687), NASA (ATP 17-ATP17-0141, 19-ATP19-0188), and STScI (HST-AR- 15018).

%
\vspace{5mm}



\bibliography{CRdissoc}{}
\bibliographystyle{aasjournal}



\end{document}